\documentclass[submission,copyright,creativecommons]{eptcs}
\providecommand{\event}{EXPRESS/SOS 2014}
\usepackage{breakurl}        
   
\usepackage{svn-multi}
\usepackage{amsmath}
\usepackage{mathtools}
\usepackage{amssymb}
\usepackage{verbatim} 
\usepackage{stmaryrd}
\usepackage{paralist}
\usepackage{cite}
\usepackage{appendix}
\usepackage{hyperref}
\usepackage[]{units}
\usepackage{tikz}
\usepackage{subfig}
\usepackage{array}
\usepackage{arydshln}
\usepackage{textcomp}
\usepackage{txfonts}
\usepackage{latexsym}
\usepackage{eufrak}
\usetikzlibrary{calc,arrows,positioning}
\allowdisplaybreaks

\svnid{$Id: fixpoint-sos.tex 2872 2014-07-26 13:51:33Z simone $}

\newif\ifdraft\drafttrue

\newcommand{\SOSrule}[2]{\frac{\displaystyle #1}{\displaystyle #2}}

\newcommand{\T}{\textsf{T}}
\newcommand{\N}{\mathbb{N}} 
\newcommand{\Ninfty}{\mathbb{N}^\infty} 
\newcommand{\Q}{\mathbb{Q}} 
\newcommand{\R}{\mathbb{R}} 
\newcommand{\Rgez}{\mathbb{R}_{\ge 0}}
\newcommand{\Rgezinfty}{\mathbb{R}_{\ge 0}^\infty} 

\newcommand{\trans}[1][]{\xrightarrow{\, {#1} \, }}
\newcommand{\ntrans}[1][]{\mathrel{{\trans[#1]}\makebox[0em][r]{$\not$\hspace{2ex}}}{\!}}

\newcommand{\rank}{r}
\newcommand{\openT}{\mathbb{T}}
\newcommand{\openST}{\mathbb{T}}
\newcommand{\openTerms}{\openSTerms \cup \openDTerms}
\newcommand{\openSTerms}{\openST(\Sigma)}
\newcommand{\closedSTerms}{\T(\Sigma)}
\newcommand{\openDT}{\mathbb{T}}
\newcommand{\openDTerms}{\openDT(\Gamma)}

\newcommand{\closedTerms}{\T(\Sigma)}
\newcommand{\VarTerm}{\mathop{\textit{Var}}}

\newcommand{\pprem}[1]{\textrm{pprem}(#1)}
\newcommand{\nprem}[1]{\textrm{nprem}(#1)}

\newcommand{\conc}[1]{\textrm{conc}(#1)}

\newcommand{\Red}[2]{\textrm{Red}}

\newcommand{\SVar}{\mathcal{V}\!_s}
\newcommand{\DVar}{\mathcal{V}\!_d}
\newcommand{\Var}{\mathcal{V}}

\newcommand{\Act}{A}
\newcommand{\tick}{{\surd}}

\newcommand{\ntyfxt}{\ensuremath{\mathit{ntyft}\mkern-1.75mu /\mkern-1.75mu \mathit{ntyxt}}}

\DeclareMathOperator{\source}{src}
\DeclareMathOperator{\target}{trgt}
\newcommand{\MVar}{\textsf{Var}}

\DeclareMathOperator{\expbound}{\Lambda}

\DeclareMathOperator{\bisimmetric}{{\bf d}}

\newcommand{\botd}{\ensuremath{\mathbf{0}}}
\newcommand{\bisimd}{\ensuremath{\mathbf{d}}}

\newcommand{\lfp}{\omega}

\newcommand{\multSym}{\ensuremath{\mathcal{M}}}
\newcommand{\multDom}{\ensuremath{\Var \to \Ninfty}}
\newcommand{\pmultSym}{\ensuremath{\mathcal{P}}}
\newcommand{\pmultDom}{\ensuremath{\Delta(\multSym)}}
\newcommand{\npmultSym}{\ensuremath{\mathcal{D}}}

\newcommand{\distSym}{\ensuremath{\mathcal{E}}}
\newcommand{\distDom}{\ensuremath{\Var \to [0,1)}}

\newcommand{\ordMultSym}{\ensuremath{\sqsubseteq}}
\newcommand{\ordPMultSym}{\ensuremath{\sqsubseteq}}
\newcommand{\ordSPMultSym}{\ensuremath{\sqsubseteq}}

\DeclareMathOperator{\dda}{\mathbf{D}}
\DeclareMathOperator{\pda}{\mathbf{P}}
\DeclareMathOperator{\da}{\mathbf{A}}
\DeclareMathOperator{\approxFunctor}{\mathbf{F}}

\DeclareMathOperator{\termda}{\tau}
\DeclareMathOperator{\ruleda}{\rho}

\DeclareMathOperator{\Kantorovich}{\mathbf{K}}
\DeclareMathOperator{\Hausdorff}{\mathbf{H}}
\DeclareMathOperator{\Bisimulation}{\mathbf{B}}

\newcommand{\nullM}{\ensuremath{0}}

\newcommand{\nullDPM}{\ensuremath{0}}

\newcommand{\downset}[1]{{\downarrow\! #1}}

\usepackage{etex}
\reserveinserts{30}

\newenvironment{apx-proof}[1] 
        {\noindent \rm \textbf{Proof of #1.}} 
        {\qed}

\newcommand{\remarkDG}[1]{}
\newcommand{\remarkST}[1]{}
\newcommand{\remarkDW}[1]{}
\newcommand{\remarkWF}[1]{}

\newcommand*\widefbox[1]{\fbox{\hspace{1em}#1\hspace{1em}}}

\newtheorem{definition}{Definition}
\newtheorem{theorem}{Theorem}

\newtheorem{proposition}{Proposition}

\newtheorem{remark}{Remark}

\newtheorem{example}{Example}
\newenvironment{proof}{{\bf Proof. }}{$\ \ \ \ \ \ \square$\\ \smallskip}
\newcommand{\qed}{\hfill\ensuremath{\square}}

\usepackage{empheq}

\title{Fixed-point Characterization of Compositionality Properties of Probabilistic Processes Combinators}

\author{Daniel Gebler
\institute{
        Department of Computer Science, VU University Amsterdam,\\
        De Boelelaan 1081a, NL-1081~HV~Amsterdam, The Netherlands}
\email{e.d.gebler@vu.nl}
\and
Simone Tini
\institute{
        Department of Scienza e Alta Tecnologia, \\
        University of Insubria, Via Valleggio 11, I-22100, Como, Italy}
\email{simone.tini@uninsubria.it}
}
\def\titlerunning{Characterization of Compositionality Properties of Probabilistic Process Combinators}
\def\authorrunning{D. Gebler \& S. Tini}

\begin{document}

\maketitle

\begin{abstract}
Bisimulation metric is a robust behavioural semantics for probabilistic processes. Given any SOS specification of probabilistic processes, we provide a method to compute for each operator of the language its respective metric compositionality property. The compositionality property of an operator is defined as its modulus of continuity which gives the relative increase of the distance between processes when they are combined by that operator. The compositionality property of an operator is computed by recursively counting how many times the combined processes are copied along their evolution. The compositionality properties allow to derive an upper bound on the distance between processes by purely inspecting the operators used to specify those processes. 
\\[1ex]{\bf Keywords:} SOS,
probabilistic transition systems,
bisimulation metric,
compositionality,
continuity
\end{abstract}

\section{Introduction}\label{sec:introduction}

Over the last decade a number of researchers have started to develop a theory of structural operational semantics for probabilistic transition systems (PTSs). Several rule formats for various PTSs were proposed that ensure compositionality of bisimilarity~\cite{Bar04,LT09,DL12} and of approximate bisimilarity~\cite{Tin10,GT13}. 
We will consider specifications with rules of the probabilistic GSOS format~\cite{Bar04,LGD12,DGL14c} in order to describe nondeterministic probabilistic transition systems~\cite{Seg95a}.

Bisimilarity is very sensitive to the exact probabilities of transitions. The slightest perturbation of the probabilities can destroy bisimilarity. Bisimulation metric~\cite{GJS90,BW06,BW05,DJGP02,DGJP04,DCPP06,DD11} provides a robust semantics for probabilistic processes. It is the quantitative analogue to bisimulation equivalence and assigns to each pair of processes a distance which measures the proximity of their quantitative properties. The distances form a pseudometric with bisimilar processes at distance $0$. 
Alternative approaches towards a robust semantics for probabilistic processes are approximate bisimulation~\cite{GJS90,DLT08,TDZ11} and bisimulation degrees~\cite{Yin02b}. 
We consider bisimulation metrics as convincingly argued in e.g.~\cite{GJS90,DGJP04,BW05}.

For compositional specification and reasoning it is necessary that the considered behavioral semantics is compatible with all operators of the language. For bisimulation metric semantics this is the notion of uniform continuity. 
Intuitively, an operator is uniformly continuous if processes composed by that operator stay close whenever their respective subprocesses are replaced by close subprocesses.

In the 1990s, rule formats that guarantee compositionality of the specified operators have been proposed by (reasonable) argumentation for admissible rules. Prominent examples are the GSOS format~\cite{BIM95} and the \ntyfxt~\cite{Gro93} format. More recently, the development of compositional proof systems for the satisfaction relation of HML-formulae~\cite{FvGdW06b,GF12} allowed to derive rule formats
from the logical characterization of the behavioral relation under investigation~\cite{BFvG04,FvGdW06a,FvGdW06c,FvGdW12}. 

We propose a new approach that allows to derive for any given specification the compositionality property of each of its specified operators. The compositionality properties are derived from an appropriate denotational model of the specified language. First, we develop for a concrete process algebra an appropriate denotational model. The denotation of an open process term describes for each resolution of the nondeterministic choices how many instances of each process variable are spawned while the process evolves. The number of spawned process replicas is weighted by the likelihood of its realization just like the bisimulation metric weights the distance between target states by their reachability.  We derive from the denotation of an open process term an upper bound on the bisimulation distance between the closed instances of the denoted process. Then we generalize this method to arbitrary processes whose operational semantics is specified by probabilistic GSOS rules.
In fact, the upper bound on the bisimulation distance between closed instances of $f(x_1,\ldots,x_{\rank(f)})$ is a modulus of continuity of operator $f$ if the denotation of $f(x_1,\ldots,x_{\rank(f)})$ is finitely bounded. 
In this case the operator $f$ is uniformly continuous and admits for compositional reasoning wrt.\ bisimulation metric.

This paper continues our research programme towards a theory of robust specifications for probabilistic processes. Earlier work~\cite{GT13} investigated compositional process combinators with respect to approximate bisimulation.
Besides the different semantics considered in this paper, we extend substantially on the approach of~\cite{GT13} by using the newly developed denotational approach. The denotational model separates clearly between nondeterministic choice, probabilistic choice, and process replication. This answers also the open question of~\cite{GT13} how the distance of processes composed by process combinators with a nondeterministic operational semantics can be approximated.
\section{Preliminaries}\label{sec:preliminaries}

\subsection{Probabilistic Transition Systems}
A \emph{signature} is a structure $\Sigma = (F, \rank)$, where 
\begin{inparaenum}[(i)]
	\item $F$ is a countable set of \emph{operators}, and
	\item $\rank \colon F \to \N$ is a \emph{rank function}.  
\end{inparaenum}
$r(f)$ gives the arity of operator $f$. We write $f\in\Sigma$ for $f\in F$. We assume an infinite set of \emph{state variables} $\SVar$ disjoint from $F$. The set of $\Sigma$-terms (also called \emph{state terms}) over $V \subseteq \SVar$, notation $T(\Sigma, V)$, is the least set satisfying: 
\begin{inparaenum}[(i)]
	\item $V \subseteq T(\Sigma, V)$, and
	\item $f(t_1, \ldots, t_{\rank(f)}) \in T(\Sigma, V)$ for $f \in \Sigma$ and $t_1, \ldots, t_{\rank(f)} \in T(\Sigma, V)$.
\end{inparaenum}
$T(\Sigma, \emptyset)$ is the set of all \emph{closed terms} and abbreviated as $\closedSTerms$. $T(\Sigma, \SVar)$ is the set of \emph{open terms} and abbreviated as $\openSTerms$. We may refer to operators as \emph{process combinators}, to variables as \emph{process variables}, and to closed terms as \emph{processes}. $\VarTerm(t)$ denotes the set of all state variables in $t$. 

Probability distributions are mappings $\pi \colon \closedSTerms \to [0,1]$ with  $\sum_{t \in \closedTerms} \pi(t) = 1$ that assign to each closed term $t \in \closedTerms$ its respective probability $\pi(t)$. By $\Delta(\closedSTerms)$ we denote the set of all probability distributions on $\closedSTerms$. We let $\pi$ range over $\Delta(\closedSTerms)$. 
The probability mass of $T \subseteq \closedSTerms$ in $\pi$ is defined by $\pi(T)=\sum_{t\in T}\pi(t)$.
Let $\delta_t$ for $t \in \closedSTerms$ denote the \emph{Dirac distribution}, i.e., $\delta_t(t)=1$ and  $\delta_t(t')=0$ if $t$ and $t'$ are syntactically not equal. The convex combination $\sum_{i \in I} q_i \pi_i$ of a family $\{\pi_i\}_{i \in I}$ of probability distributions $\pi_i \in \Delta(\closedSTerms)$ with $q_i \in (0,1]$ and $\sum_{i \in I} q_i = 1$ is defined by $(\sum_{i \in I}q_i \pi_i)(t) = \sum_{i \in I} (q_i \pi_i(t))$. 
By $f(\pi_1,\dots,\pi_{\rank(f)})$ we denote the distribution defined by $f(\pi_1,\dots,\pi_{\rank(f)})(f(t_1,\ldots,t_{\rank(f)}))$ = $\prod_{i=1}^{\rank(f)}\pi_i(t_i)$. 
We may write $\pi_1 \, f \, \pi_2$ for $f(\pi_1,\pi_2)$.

In order to describe probabilistic behavior, we need expressions that denote probability distributions. We assume an infinite set of distribution variables $\DVar$. We let $\mu$ range over $\DVar$, and $x, y$ range over $\Var = \SVar \cup \DVar$. The set of \emph{distribution terms} over state variables $V_s \subseteq \SVar$ and distribution variables $V_d \subseteq \DVar$, notation $T(\Gamma, V_s, V_d)$ with $\Gamma$ denoting the signature extending $\Sigma$ by operators to describe distributions, is the least set satisfying: 
\begin{inparaenum}[(i)]
	\item \label{def:DT:var_and_inst_dirac}
		$V_d \cup \{\delta(t) \mid t \in \T(\Sigma, V_s)\} \subseteq \T(\Gamma, V_s, V_d)$, 
	\item \label{def:DT:sum} 
		${\textstyle \sum_{i\in I} q_i \theta_i \in \T(\Gamma, V_s, V_d)}$ if $\theta_i \in \T(\Gamma, V_s, V_d)$ and $q_i \in (0,1]$ with $\sum_{i\in I} q_i = 1$, and
	\item \label{def:DT:prod} 
		$f(\theta_1,\ldots,\theta_{\rank(f)}) \in \T(\Gamma, V_s, V_d)$ if $f \in \Sigma$ and $\theta_i \in \T(\Gamma, V_s, V_d)$.
\end{inparaenum}
A \emph{distribution variable} $\mu \in \DVar$ is a variable that takes values from $\Delta(\closedSTerms)$. An \emph{instantiable Dirac distribution} $\delta(t)$ is an expression that takes as value the Dirac distribution $\delta_{t'}$ when variables in $t$ are substituted so that $t$ becomes the closed term $t'$. Case (\ref{def:DT:sum}) allows to construct convex combinations of distributions. 
We write $\theta_1 \oplus_q \theta_2$ for $q\theta_1 + (1-q)\theta_2$.
Case (\ref{def:DT:prod}) lifts the structural inductive construction of state terms to distribution terms. 
$\openDTerms$ denotes $T(\Gamma, \SVar,\DVar)$.
$\VarTerm(\theta)$ denotes the set of all state and distribution variables in $\theta$.

A \emph{substitution} is a mapping $\sigma \colon \Var \to \openTerms$ such that $\sigma(x) \in \openSTerms$ if $x\in \SVar$, and $\sigma(\mu) \in \openDTerms$ if $\mu\in \DVar$. A substitution extends to a mapping from state terms to state terms as usual.
A substitution extends to distribution terms by $\sigma(\delta(t))=\delta_{\sigma(t)}$, $\sigma(\sum_{i\in I} q_i \theta_i) = \sum_{i\in I} q_i \sigma(\theta_i)$ and $\sigma(f(\theta_1,\ldots,\theta_{\rank(f)})) = f(\sigma(\theta_1),\ldots,\sigma(\theta_{\rank(f)}))$. Notice that closed instances of distribution terms are probability distributions. 

Probabilistic transition systems generalize labelled transition systems (LTSs) by allowing for probabilistic choices in the transitions. We consider nondeterministic probabilistic LTSs (Segala-type systems)~\cite{Seg95a} with countable state spaces.

\begin{definition}[PTS]
A \emph{nondeterministic probabilistic labeled transition system} (PTS) is given by a triple $(\closedSTerms,\Act,{\trans})$, where $\Sigma$ is a signature, $\Act$ is a countable set of \emph{actions}, and ${\trans} \subseteq {\closedSTerms \times \Act \times \Delta(\closedSTerms)}$ is a \emph{transition relation}.
\end{definition}
We write $t \trans[a] \pi$ for ${(t,a,\pi)} \in {\trans}$, and $t \trans[a]$ if $t \trans[a] \pi$ for some $\pi \in \Delta(\closedSTerms)$.

\subsection{Specification of Probabilistic Transition Systems}

We specify PTSs by SOS rules of the probabilistic GSOS format~\cite{Bar04} and adapt from~\cite{LGD12} the language to describe distributions. We do not consider quantitative premises 
because they are incompatible\footnote{Cases~8 and~9 
in~\cite{GT13} show that rules with quantitative premises may define operators that are not compositional wrt. approximate bisimilarity. The same holds for metric bisimilarity.
} with compositional approximate reasoning.

\begin{definition}[PGSOS rule]\label{def:pgsos}
A \emph{PGSOS rule} has the form:
\[
	\SOSrule{\{ x_i \trans[a_{i,m}] \mu_{i,m} \mid i \in I, m \in M_i \} \qquad
			 \{ x_i \ntrans[b_{i,n}] \mid i \in I, n \in N_i \}}
			{ f(x_1,\ldots,x_{\rank(f)}) \trans[a] \theta}
\]
with $I = \{1,\dots,\rank(f)\}$ the indices of the arguments of operator $f \in \Sigma$, finite index sets $M_i,N_i$, actions $a_{i,m},b_{i,n},a \in \Act$, state variables $x_i \in \SVar$, distribution variables $\mu_{i,m} \in \DVar$,  distribution term $\theta\in\openDTerms$, and constraints:
\begin{enumerate}
	\item all $\mu_{i,m}$ for $i \in I, m \in M_i$ are pairwise different; 
	\item all $x_1,\ldots,x_{\rank(f)}$ are pairwise different;
	\item $\VarTerm(\theta) \subseteq \{\mu_{i,m} \mid i\in I, m\in M_i \} \cup \{ x_1 \ldots,x_{\rank(f)}\}$.
\end{enumerate}
\end{definition}
The expressions $x_i \trans[a_{i,m}] \mu_{i,m}$ (resp.\ $x_i \ntrans[b_{i,n}]$) above the line are called \emph{positive} (resp. \emph{negative}) \emph{premises}. We call $\mu_{i,m}$ in $x_i \trans[a_{i,m}] \mu_{i,m}$ a \emph{derivative} of $x_i$. We denote the set of positive (resp.\ negative) premises of rule $r$ by $\pprem{r}$ (resp.\ $\nprem{r}$). 
The expression $f(x_1,\dots,x_{\rank(f)}) \trans[a] \theta$ below the line is called the \emph{conclusion}, notation $\conc{r}$, $f(x_1,\dots,x_{\rank(f)})$ is called the \emph{source}, notation $\source(r)$, the $x_i$ are called the \emph{source variables}, notation $x_i \in \source(r)$, and $\theta$ is called the \emph{target}, notation $\target(r)$. 

A \emph{probabilistic transition system specification} (PTSS) in PGSOS format is a triple $P = (\Sigma, \Act, R)$, where $\Sigma$ is a signature, $\Act$ is a countable set of actions and $R$ is a countable set of PGSOS rules. 
$R_f$ is the set of those rules of $R$ with source $f(x_1,\dots,x_{\rank(f)})$. A \emph{supported model} of $P$ is a PTS $(\closedSTerms, \Act, {\trans})$ such that $t\trans[a] \pi \in {\trans}$ iff for some rule $r \in R$ and some closed substitution $\sigma$  all premises of $r$ hold, 
i.e. for all $x_i \trans[a_{i,m}] \mu_{i,m} \in \pprem{r}$ we have $\sigma(x_i) \trans[a_{i,m}] \sigma(\mu_{i,m}) \in {\trans}$ and for all $x_i \ntrans[b_{i,n}] \in \nprem{r}$ we have $\sigma(x_i) \trans[b_{i,n}] \pi \not\in {\trans}$ for all $\pi \in \Delta(\closedSTerms)$, and the conclusion $\conc{r} = f(x_1,\ldots,x_{\rank(f)}) \trans[a] \theta$ instantiates to $\sigma(f(x_1,\ldots,x_{\rank(f)})) = t$ and $\sigma(\theta) = \pi$. 
Each PGSOS PTSS has exactly one supported model~\cite{BIM95,Bar02} which is moreover finitely branching.

\subsection{Bisimulation metric on Probabilistic Transition Systems}
Behavioral pseudometrics are the quantitative analogue to behavioral equivalences and formalize the notion of \emph{behavioral distance} between processes. A \emph{$1$-bounded pseudometric} is a function $d \colon \closedSTerms \times \closedSTerms \to [0,1]$  with
\begin{inparaenum}[(i)]
	\item $d(t,t)= 0$,
	\item $d(t,t') = d(t',t)$, and
	\item $d(t,t') \le d(t,t'') + d(t'',t')$,
\end{inparaenum}
for all terms $t,t',t'' \in \closedSTerms$.

We define now bisimulation metrics as quantitative analogue to bisimulation equivalences. Like for 
bisimulation we need to lift the behavioral pseudometric on states $\closedSTerms$ to distributions $\Delta(\closedSTerms)$ and sets of distributions $P(\Delta(\closedSTerms))$. 
A \emph{matching} $\omega \in \Delta(\closedSTerms \times \closedSTerms)$ for $(\pi,\pi') \in \Delta(\closedSTerms) \times \Delta(\closedSTerms)$ is given if $\sum_{t'\in \closedSTerms} \omega(t,t')=\pi(t)$ and $\sum_{t\in \closedSTerms} \omega(t,t')=\pi'(t')$ for all $t,t'\in \closedSTerms$. We denote the set of all matchings for $(\pi,\pi')$ by $\Omega(\pi,\pi')$. 
The \emph{Kantorovich pseudometric} $\Kantorovich(d)\colon \Delta(\closedSTerms) \times \Delta(\closedSTerms) \to [0,1]$ is defined for a pseudometric $d\colon \closedSTerms \times \closedSTerms \to [0,1]$ by 
\[
	\Kantorovich(d)(\pi,\pi') = \min_{\omega \in \Omega(\pi,\pi')} \sum_{t,t'\in \closedSTerms}d(t,t') \cdot \omega(t,t')
\]
for $\pi,\pi' \in \Delta(\closedSTerms)$. 
%
The \emph{Hausdorff pseudometric} $\Hausdorff(\hat{d})\colon P(\Delta(\closedSTerms)) \times P(\Delta(\closedSTerms)) \to [0,1]$ is defined for a pseudometric $\hat{d}\colon \Delta(\closedSTerms) \times \Delta(\closedSTerms) \to [0,1]$ by 
\[
	\Hausdorff(\hat{d})(\Pi_1,\Pi_2) = \max \left\{ \adjustlimits\sup_{\pi_1 \in \Pi_1}\inf_{\pi_2 \in \Pi_2} \hat{d}(\pi_1,\pi_2), \adjustlimits\sup_{\pi_2\in \Pi_2}\inf_{\pi_1\in \Pi_1} \hat{d}(\pi_2,\pi_1) \right\}
\]
for $\Pi_1,\Pi_2 \subseteq \Delta(\closedSTerms)$ whereby $\inf \emptyset = 1$ and $\sup \emptyset = 0$.

A bisimulation metric is a pseudometric on states such that for two states each transition from one state can be mimicked by a transition from the other state and the distance between the target distributions does not exceed the distance of the source states. 

\begin{definition}[Bisimulation metric] \label{def:bisim_metric}
A $1$-bounded pseudometric $d$ on $\closedSTerms$ is a \emph{bisimulation metric} if for all $t,t'\in \closedSTerms$ with $d(t,t')<1$, if $t \trans[a] \pi$ then there exists a transition $t' \trans[a] \pi'$ with $\Kantorovich(d)(\pi,\pi') \le d(t,t')$.
\end{definition}
We order bisimulation metrics $d_1 \sqsubseteq d_2$ iff $d_1(t,t') \le d_2(t,t')$ for all $t,t' \in \closedSTerms$. The smallest bisimulation metric, notation $\bisimd$, is called \emph{bisimilarity metric} and assigns to each pair of processes the least possible distance. We call the bisimilarity metric distance also bisimulation distance. Bisimilarity equivalence~\cite{LS91,Seg95a} is the kernel of the bisimilarity metric~\cite{DCPP06}, i.e. $\bisimd(t,t') = 0$ iff $t$ and $t'$ are bisimilar. We say that processes $t$ and $t'$ \emph{do not totally disagree} if $\bisimd(t,t') < 1$. 

\begin{remark}\label{rem:agreement_immidiate_actions}
Let $t,t'$ be processes that do not totally disagree. Then $t \trans[a]$ iff $t' \trans[a]$ for all $a \in \Act$, i.e. $t$ and $t'$ agree on the actions they can perform immediately.
\end{remark}


Bisimulation metrics can alternatively be defined as prefixed points of a monotone function. Let $([0,1]^{\closedSTerms \times \closedSTerms},\sqsubseteq)$ be the complete lattice defined by  $d \sqsubseteq d'$ iff $d(t, t') \le d'(t, t')$, for all $t, t' \in \closedSTerms$. 
We define the function $\Bisimulation \colon [0,1]^{\closedSTerms \times \closedSTerms} \to [0,1]^{\closedSTerms \times \closedSTerms}$ for $d \colon \closedSTerms \times \closedSTerms \to [0,1]$ and $t,t' \in \closedSTerms$ by:
\[
	\Bisimulation(d)(t,t') = \sup_{a\in A} \left\{ \Hausdorff(\Kantorovich(d))(\mathit{der}(t,a), \mathit{der}(t',a)) \right\}
\]
with $\mathit{der}(t,a) = \{\pi \mid t \trans[a] \pi \}$.
%
%
\begin{proposition}[\!\!\protect{\cite{DCPP06}}] \label{prop:bisim_metric_lfp_D}
The bisimilarity metric $\bisimd$ is the least fixed point of $\Bisimulation$.
\end{proposition}

\section{Denotational model} \label{sec:denotational_model_PA}

We develop now a denotational model for open terms. Essentially, the denotation of an open term $t$ describes for each variable in $t$ how many copies are spawned while $t$ evolves. The denotation of $t$ allows us to formulate an upper bound on the bisimulation distance between closed instances of $t$. In this section we consider a concrete process algebra. In the next section we generalize our method to arbitrary PGSOS specifications.

Let $\Sigma_{\text{PA}}$ be the signature of the core operators of the probabilistic process algebra in~\cite{DL12} defined by the stop process $0$, a family of $n$-ary prefix operators $a.([q_1]\_ \oplus \cdots \oplus [q_n]\_ )$ with $a \in \Act$, $n\ge 1$, $q_1,\ldots,q_n \in (0,1]$ and $\sum_{i=1}^n q_i = 1$, alternative composition $\_ + \_$,
and parallel composition $\_ \parallel_B \_$ for each $B \subseteq \Act$. 
We write $a.\bigoplus_{i=1}^n[q_i]\_$ for $a.([q_1]\_ \oplus \cdots \oplus [q_n]\_ )$, and $a.\_$ for $a.[1]\_$ (deterministic prefix operator). Moreover, we write $\_ \parallel \_$ \ for \ $\_ \parallel_\Act \_$ (synchronous parallel composition). The PTSS $P_{\text{PA}} = (\Sigma_{\text{PA}}, \Act, R_{\text{PA}})$ is given by the following PGSOS rules in $R_{\text{PA}}$:

\vspace{0.25cm}
\begin{tabular}{ c@{\hskip 1.3cm} c@{\hskip 1.3cm} c@{\hskip 1.3cm} }
	$\displaystyle\SOSrule{}{a.\bigoplus_{i=1}^n[q_i]x_i \trans[a] \sum_{i=1}^n q_i \delta(x_i)}$
&
	$\displaystyle\SOSrule{x_1 \trans[a] \mu_1}{x_1 + x_2 \trans[a] \mu_1}$
&
	$\displaystyle\SOSrule{x_2 \trans[a] \mu_2}{x_1 + x_2 \trans[a] \mu_2}$ \\[1.25cm]
    $\displaystyle\SOSrule{x_1\trans[a]\mu_1 \quad x_2\trans[a]\mu_2 \quad (a \in B)}{{x_1 \parallel_B x_2} \trans[a] {\mu_1 \parallel_B \mu_2}}$
&
                $\displaystyle\SOSrule{x_1\trans[a]\mu_1 \quad (a \not\in B)}{{x_1 \parallel_B x_2} \trans[a] {\mu_1 \parallel_B \delta(x_2)}}$
&
	$\displaystyle\SOSrule{x_2\trans[a]\mu_2 \quad (a \not\in B)}{{x_1 \parallel_B x_2} \trans[a] {\delta(x_1) \parallel_B \mu_2}}$\\
\end{tabular}
\vspace{0.5cm}


We call the open terms $\openT(\Sigma_{\text{PA}})$ \emph{nondeterministic probabilistic process terms}. We define two important subclasses of $\openT(\Sigma_{\text{PA}})$ that allow for a simpler approximation of the distance of their closed instances. Let $\openT_{\text{det}}(\Sigma_{\text{PA}})$ be the set of \emph{deterministic process terms}, which are those terms of $\openT(\Sigma_{\text{PA}})$ that are built exclusively from the stop process $0$, deterministic prefix $a.\_$, and synchronous parallel composition $\_ \parallel \_$ (no nondeterministic and no probabilistic choices).
We call the open terms $\openT_{\text{det}}(\Sigma_{\text{PA}})$ deterministic because all probabilistic or nondeterministic choices in the operational semantics of the closed instances $\sigma(t)$, with $\sigma \colon \SVar \to \T(\Sigma_{\text{PA}})$ any closed substitution, arise exclusively from the processes in $\sigma$.
Let $\openT_{\text{prob}}(\Sigma_{\text{PA}})$ be the set of \emph{probabilistic process terms}, which are those terms of $\openT(\Sigma_{\text{PA}})$ that are built exclusively from the stop process $0$, probabilistic prefix $a.\bigoplus_{i=1}^n[q_i]\_$, and synchronous parallel composition $\_ \parallel \_$ (no nondeterministic choices).
Again, all nondeterministic choices in $\sigma(t)$ arise exclusively from the processes in $\sigma$.


The denotation of a deterministic process term $t \in \openT_{\text{det}}(\Sigma_{\text{PA}})$ is a mapping $m \colon \multDom$ that describes for each process variable $x \in \VarTerm(t)$ how many copies of $x$ or some derivative of $x$ are spawned while $t$ evolves. We call $m$ the \emph{multiplicity} of $t$. Let $\multSym$ be the set of all mappings $\multDom$. The denotation of $t$, notation $\llbracket t \rrbracket_\multSym$,
is defined by 
$\llbracket 0 \rrbracket_\multSym (x) = 0$, 
$\llbracket x \rrbracket_\multSym (x) = 1$, 
$\llbracket x \rrbracket_\multSym (y) = 0$ if $x \neq y$, 
$\llbracket t_1 \parallel t_2 \rrbracket_\multSym (x) = \llbracket t_1 \rrbracket_\multSym (x) + \llbracket t_2 \rrbracket_\multSym (x)$, and
$\llbracket a.t' \rrbracket_\multSym (x) = \llbracket t' \rrbracket_\multSym (x)$. 

We use notation $\nullM \in \multSym$ for the multiplicity that assigns $0$ to each $x \in \Var$, and $n_V \in \multSym$ with $V \subseteq \Var$ for the multiplicity such that $n_V(x) = n$ if $x \in V$ and $n_V(x) = 0$ if $x \not\in V$. We write $n_{x}$ for $n_{\{x\}}$. 
As it will become clear in the next sections, we need the denotation $m(x)=\infty$ for (unbounded) recursion and replication.

We will approximate the bisimulation distance between $\sigma_1(t)$ and $\sigma_2(t)$ for closed substitutions $\sigma_1,\sigma_2$ using the denotation of $t$ and the bisimulation distances between processes $\sigma_1(x)$ and $\sigma_2(x)$ of variables $x \in \VarTerm(t)$. The bisimulation distance of variables is represented by a mapping $e \colon \distDom$. We call $e$ a \emph{process distance}. Let $\distSym$ be the set of all process distances \distDom. 
We henceforth assume closed substitutions $\sigma_1,\sigma_2$ with a bisimulation distance between $\sigma_1(x)$ and $\sigma_2(x)$ that is strictly less than $1$. Practically, this is a very mild restriction because for any (non-trivial) process combinator the composition of processes that totally disagree (i.e. which are in bisimulation distance $1$) may lead to composed processes that again totally disagree. 
For any $d \colon \closedSTerms \times \closedSTerms \to [0,1]$ and any closed substitutions $\sigma_1,\sigma_2$ we define the associated process distance $d(\sigma_1,\sigma_2) \in \distSym$ by $d(\sigma_1,\sigma_2)(x) = d(\sigma_1(x),\sigma_2(x))$. 

\begin{definition}
For a multiplicity $m \in \multSym$ and process distance $e \in \distSym$ we define the \emph{deterministic distance approximation from above} as
\[
	\dda(m,e) = 1 - \prod_{x \in \Var} (1-e(x))^{m(x)}
\]
\end{definition}
To understand the functional $\dda$ remind that $e(x)$ is the distance between processes $\sigma_1(x)$ and $\sigma_2(x)$. In other words, processes $\sigma_1(x)$ and $\sigma_2(x)$ disagree by $e(x)$ on their behavior. Hence, $\sigma_1(x)$ and $\sigma_2(x)$ agree by $1-e(x)$. Thus, $m(x)$ copies of $\sigma_1(x)$ and $m(x)$ copies of $\sigma_2(x)$ agree by at least $\prod_{x \in \Var}(1-e(x))^{m(x)}$, and disagree by at most $1-\prod_{x \in \Var}(1-e(x))^{m(x)}$.

\begin{example} \label{ex:det_distance_approx}
Consider the deterministic process term $t = x \parallel x$ and substitutions $\sigma_1(x)=a.a.0$ and $\sigma_2(x)=a.([0.9]a.0 \oplus [0.1]0)$. In this and all following examples we assume that $\sigma_1$ and $\sigma_2$ coincide on all other variables for which the substitution is not explicitly defined, i.e.\  $\sigma_1(y)=\sigma_2(y)$ if $x \neq y$ in this example.
It is clear that $\bisimd(\sigma_1(x),\sigma_2(x)) = 0.1$.  
Then, $\bisimd(\sigma_1(t),\sigma_2(t))=0.1\cdot 0.9 + 0.9 \cdot 0.1 + 0.1 \cdot 0.1 = 0.19$, which is the likelihood that either the first, the second or both arguments of 
$\sigma_2(x \parallel x)$ can perform action $a$ only once. The denotation of $t$ is $\llbracket t \rrbracket_\multSym (x) = 2$. Then, $\dda(\llbracket t \rrbracket_\multSym,\bisimd(\sigma_1,\sigma_2))=1-(1-0.1)^2=0.19$. 
\end{example}

The functional $\dda$ defines an upper bound on the bisimulation distance of deterministic processes.

\begin{proposition} \label{prop:upper_bound_dda}
Let $t \in \openT_{\text{det}}(\Sigma_{\text{PA}})$ be a deterministic process term and $\sigma_1,\sigma_2$ be closed substitutions. Then $\bisimd(\sigma_1(t),\sigma_2(t)) \le \dda(\llbracket t \rrbracket_\multSym,\bisimd(\sigma_1,\sigma_2))$. 
\end{proposition}

The distance $\bisimd(\sigma_1,\sigma_2)$ abstracts from the concrete reactive behavior of terms $\sigma_1(x)$ and $\sigma_2(x)$. It is not hard to see that for deterministic process terms without parallel composition the approximation functional $\dda$ gives the exact bisimulation distance. However, the parallel composition of processes may lead to an overapproximation if the bisimulation distance of process instances arises (at least partially) from reactive behavior on which the processes cannot synchronize.

\begin{example}\label{prop:over_upper_bound_dda}
Consider $t = x \parallel a.a.0$ and substitutions $\sigma_1(x)=a.b.0$ and $\sigma_2(x)=a.([0.9]b.0 \oplus [0.1]0)$ with $\bisimd(\sigma_1(x),\sigma_2(x)) = 0.1$. We have $\bisimd(\sigma_1(t),\sigma_2(t))=0$ since both $\sigma_1(t)$ and $\sigma_2(t)$ make an $a$ move to a distribution of parallel compositions either $b.0 \parallel a.0$ or $0 \parallel a.0$ that all cannot proceed. Note that the bisimulation distance between $\sigma_1(x)$ and $\sigma_2(x)$ arises from the difference on performing action $b$ which cannot synchronize with $a$. The denotation of $t$ is $\llbracket t \rrbracket_\multSym (x) = 1$ which gives in this case an overapproximation of the distance $\bisimd(\sigma_1(t),\sigma_2(t))=0 < \dda(\llbracket t \rrbracket_\multSym,\bisimd(\sigma_1,\sigma_2))=1-(1-0.1)=0.1$. However, for $\sigma_1'(x)=a.a.0$ and $\sigma_2'(x)=a.([0.9]a.0 \oplus [0.1]0)$ with $\bisimd(\sigma_1'(x),\sigma_2'(x)) = 0.1$ we get $\bisimd(\sigma_1'(t),\sigma_2'(t)) = 0.1= \dda(\llbracket t \rrbracket_\multSym,\bisimd(\sigma_1',\sigma_2'))$.
\end{example}

We remark that the abstraction of the closed substitutions to process distances is intentional and very much in line with common compositionality criteria that relate the distance of composed processes with the distance of the process components.

The denotation of a probabilistic process term $t \in \openT_{\text{prob}}(\Sigma_{\text{PA}})$ is a distribution $p \in \pmultDom$ that describes for each multiplicity $m \in \multSym$ the likelihood $p(m)$ that for each process variable $x \in \VarTerm(t)$ exactly $m(x)$ copies of $x$ or some derivative of $x$ are spawned while $t$ evolves. We call $p$ the \emph{probabilistic multiplicity} of $t$. Let $\pmultSym$ be the set of all distributions $\pmultDom$. The denotation of $t$, notation $\llbracket t \rrbracket_\pmultSym$, is defined by
$\llbracket 0 \rrbracket_\pmultSym = \delta_m$ with $m=0$,
$\llbracket x \rrbracket_\pmultSym = \delta_m$ with $m=1_x$,
$\llbracket t_1 \parallel t_2 \rrbracket_\pmultSym (m) = \sum_{{m_1,m_2 \in \multSym \atop m(x)=m_1(x)+m_2(x)} \atop \text{for all }x\in\Var} \llbracket t_1 \rrbracket_\pmultSym (m_1) \cdot \llbracket t_2 \rrbracket_\pmultSym (m_2)$, and
$\llbracket a.\bigoplus_{i=1}^n[q_i] t_i \rrbracket_\pmultSym = \sum_{i=1}^{n}q_i \llbracket t_i \rrbracket_\pmultSym$.
Notice that $\llbracket t \rrbracket_\pmultSym = \delta_{\llbracket t \rrbracket_\multSym}$ for all $t \in \openT_{\text{det}}(\Sigma_{\text{PA}})$.

For important probabilistic multiplicities we use the same symbols as for multiplicities but it will always be clear from the context if we refer to probabilistic multiplicities or multiplicities. By $0 \in \pmultSym$ we mean the probabilistic multiplicity that gives probability $1$ to the multiplicity $0 \in \multSym$. By $n_V \in \pmultSym$ we mean the probabilistic multiplicity that gives probability $1$ to the multiplicity $n_V \in \multSym$.

\begin{definition} 
For a probabilistic multiplicity $p \in \pmultSym$ and process distance $e \in \distSym$ we define the \emph{probabilistic distance approximation from above} as
\[
	\pda(p,e) = \sum_{m \in \multSym} p(m) \cdot \dda(m,e) 
\]
\end{definition}

\begin{example}
Consider $t=a.([0.5] (x \parallel x) \oplus [0.5] 0)$ and substitutions $\sigma_1(x)=a.a.0$ and $\sigma_2(x)=a.([0.9]a.0 \oplus [0.1]0)$ with $\bisimd(\sigma_1(x),\sigma_2(x)) = 0.1$. It holds that $\bisimd(\sigma_1(t),\sigma_2(t))=0.5(1-(1-0.1)^2)$. The probabilistic multiplicity of $t$ is $\llbracket t \rrbracket_\pmultSym (2_x)=0.5$ and $\llbracket t \rrbracket_\pmultSym (0)=0.5$. Then, $\dda(2_x,\bisimd(\sigma_1,\sigma_2))=1-(1-0.1)^2$ and $\dda(\nullM,\bisimd(\sigma_1,\sigma_2))=0$. Hence, we get the probabilistic distance approximation $\pda(\llbracket t \rrbracket_\pmultSym, \bisimd(\sigma_1,\sigma_2))=0.5(1-(1-0.1)^2)$. 
\end{example}

\begin{remark} \label{rem:weighting_prob_choice_replication_bernoulli}
The functional $\pda$ shows a very important interaction between probabilistic choice and process replication. Consider again the process term $t=a.([0.5] (x \parallel x) \oplus [0.5] 0)$ and any closed substitutions $\sigma_1,\sigma_2$ with $\bisimd(\sigma_1(x),\sigma_2(x))=\epsilon$ for any $\epsilon \in [0,1)$. In the probabilistic distance approximation $\pda(\llbracket t \rrbracket_\pmultSym, \bisimd(\sigma_1,\sigma_2))$ the deterministic  distance approximation $\dda(2_x,\bisimd(\sigma_1,\sigma_2))=1-(1-\epsilon)^2$ of the synchronous parallel execution $x \parallel x$ of two instances of $x$ is weighted by the likelihood $0.5$ of its realization. Hence, $\pda(\llbracket t \rrbracket_\pmultSym, \bisimd(\sigma_1,\sigma_2))=0.5(1-(1-\epsilon)^2)$. From Bernoulli's inequality $\frac{1}{m}(1-(1-\epsilon)^n) \le \epsilon$ if $m \ge n$, we get $0.5(1-(1-\epsilon)^2) \le \epsilon$. Hence, the distance between instances of two copies running synchronously in parallel with a probability of $0.5$ is at most the distance between those instances running (non-replicated) with a probability of $1.0$.
\end{remark}

\noindent
Notice that $\pda(\llbracket t \rrbracket_\pmultSym,\bisimd(\sigma_1,\sigma_2)) = \dda(\llbracket t \rrbracket_\multSym,\bisimd(\sigma_1,\sigma_2))$ for all $t \in \openT_{\text{det}}(\Sigma_{\text{PA}})$. The functional $\pda$ defines an upper bound on the bisimulation distance of probabilistic processes.

\begin{proposition} \label{prop:upper_bound_pda}
Let $t \in \openT_{\text{prob}}(\Sigma_{\text{PA}})$ be a probabilistic process term and $\sigma_1,\sigma_2$ be closed substitutions. Then $\bisimd(\sigma_1(t),\sigma_2(t)) \le \pda(\llbracket t \rrbracket_\pmultSym, \bisimd(\sigma_1,\sigma_2))$. 
\end{proposition}

Before we can introduce the denotation of nondeterministic probabilistic processes, we need to order the denotation of probabilistic processes. Let $\pi \colon \multSym \to [0,1]$ with $\sum_{m \in \multSym} \pi(m) \le 1$ be a subdistribution over multiplicities. We define the weighting of $\pi$ as a mapping $\overline{\pi}\colon \Var \to \Rgez$ defined $\overline{\pi}(x) = (1/|\pi|)\sum_{m \in \multSym} \pi(m) \cdot m(x)$ if $|\pi|>0$, with $|\pi|= \sum_{m \in \multSym} \pi(m)$ the size of $\pi$, and $\overline{\pi}(x) = 0$ if $|\pi|=0$. Intuitively, the number of process copies $m(x)$ are weighted by the probability $\pi(m)$ of realization of that multiplicity. We order probabilistic multiplicities $p_1 \sqsubseteq p_2$ if $p_1$ can be decomposed into subdistributions such that each multiplicity in $p_2$ is above some weighted subdistribution of $p_1$. The order is now defined by:
\begin{align*}
p_1 \ordPMultSym p_2 & \text{ iff there is a } \omega \in \Omega(p_1,p_2) \text{ with } \overline{\omega(\cdot,m)} \sqsubseteq m \text{ for all }m \in \multSym \\ 
	m_1 \ordMultSym m_2 & \text{ iff } m_1(x) \le m_2(x) \text{ for all } x \in \Var
\end{align*}

The denotation of a nondeterministic probabilistic process term $t \in \openT(\Sigma_{\text{PA}})$ is a set of probabilistic multiplicities $P \subseteq \pmultSym$ that describes by 
$p \in P$ some resolution of the nondeterministic choices in $t$ such that the process evolves as a probabilistic process described by $p$. 
We construct a Hoare powerdomain over the probabilistic multiplicities $\pmultSym$ and use as canonical representation for any set of probabilistic multiplicities $P \subseteq \pmultSym$ the 
downward closure defined as ${\downset{P}} = {\{ p \in \pmultSym \mid p \sqsubseteq p' \text{ for some } p' \in P\}}$.
Let $\npmultSym$ be the set of non-empty downward closed sets of probabilistic multiplicities $\{ P \subseteq \pmultSym \mid P \neq \emptyset \text{ and }\downset{P} = P \}$. 
We use downward closed sets such that $\npmultSym$ will form a complete lattice with the order defined below (esp.\ satisfies antisymmetry, cf.\ Proposition~\ref{prop:preorder_D}). The denotation of $t$, notation $\llbracket t \rrbracket$, is defined by
$\llbracket 0 \rrbracket = \{ \llbracket 0 \rrbracket_\pmultSym \}$,
$\llbracket x \rrbracket = \downset\{ \llbracket x \rrbracket_\pmultSym \}$,
$p \in \llbracket t_1 \parallel_B t_2 \rrbracket$ iff there are $p_1 \in \llbracket t_1 \rrbracket$ and $p_2 \in \llbracket t_2 \rrbracket$ such that $p \sqsubseteq p'$ with $p'$ defined by $p'(m) = \sum_{{m_1,m_2 \in \multSym \atop m(x)=m_1(x)+m_2(x)} \atop \text{for all }x\in\Var} p_1(m_1) \cdot p_2(m_2)$ for all $m \in \multSym$,
$p \in \llbracket a.\bigoplus_{i=1}^n[q_i] t_i \rrbracket$ iff there are $p_i \in  \llbracket t_i \rrbracket$ such that $p \sqsubseteq p'$ with $p'$ defined by $p'=\sum_{i=1}^{n}q_i \cdot p_i$, and 
$\llbracket t_1 + t_2 \rrbracket = \llbracket t_1 \rrbracket \cup \llbracket t_2 \rrbracket$. 
Notice that ${\llbracket t \rrbracket} = {\downset{\{\llbracket t \rrbracket_\pmultSym \}}}$ for all $t \in \openT_{\text{prob}}(\Sigma_{\text{PA}})$. 
By $0 \in \npmultSym$ we mean the singleton set containing the probabilistic multiplicity $0 \in \pmultSym$, 
and by $n_V \in \npmultSym$ the downward closure of the singleton set with element $n_V \in \pmultSym$.

\begin{definition}
For a nondeterministic probabilistic multiplicity $P \in \npmultSym$ and process distance $e \in \distSym$ we define the \emph{nondeterministic probabilistic distance approximation from above} as
\[
	\da(P,e) = \sup_{p \in P} \pda(p,e) 
\]
\end{definition}

\begin{example}
Consider the nondeterministic probabilistic process term $t=a.([0.5] (x \parallel x) \oplus [0.5] 0) + b.y$, and substitutions $\sigma_1(x)= a.a.0$, $\sigma_2(x)=a.([0.9]a.0 \oplus [0.1]0)$ and $\sigma_1(y)= b.b.0$, $\sigma_2(y)=b.([0.8]b.0 \oplus [0.2]0)$. It is clear that $\bisimd(\sigma_1(x),\sigma_2(x)) = 0.1$ and $\bisimd(\sigma_1(y),\sigma_2(y)) = 0.2$. Now, $\bisimd(\sigma_1(t),\sigma_2(t))=\max\{0.5(1-(1-0.1)^2),0.2\}$. The nondeterministic probabilistic multiplicity of $t$ is $\llbracket t \rrbracket = \downset{\{p_1,p_2\}}$, for $p_1(2_x)=0.5, p_1(\nullM)=0.5$ and $p_2(1_y) = 1.0$. Thus 
$\da(\llbracket t \rrbracket,\bisimd(\sigma_1,\sigma_2))= \max(\pda(p_1,\bisimd(\sigma_1,\sigma_2)),\pda(p_2,\bisimd(\sigma_1,\sigma_2))) = \max(0.5(1-(1-0.1)^2),0.2)$.
\end{example}

\noindent
Notice that $\da(\llbracket t \rrbracket,\bisimd(\sigma_1,\sigma_2)) = \pda(\llbracket t \rrbracket_\pmultSym,\bisimd(\sigma_1,\sigma_2))$ for all $t \in \openT_{\text{prob}}(\Sigma_{\text{PA}})$. Moreover, $\da(P,e) = \da(\downset{P},e)$ for any $P \subseteq \pmultSym$. The functional $\da$ defines an upper bound on the bisimulation distance of nondeterministic probabilistic process terms.

\begin{theorem} \label{thm:upper_bound_da}
Let $t \in \openT(\Sigma_{\text{PA}})$ be a nondeterministic probabilistic process term and $\sigma_1,\sigma_2$ be closed substitutions. Then $\bisimd(\sigma_1(t),\sigma_2(t)) \le \da(\llbracket t \rrbracket,\bisimd(\sigma_1,\sigma_2))$. 
\end{theorem}
Theorem~\ref{thm:upper_bound_da} shows that the denotation of a process term is adequate to define an upper bound on the distance of closed instances of that process term. The converse notion is full-abstraction in the sense that $\bisimd(\sigma_1(t),\sigma_2(t)) = \da(\llbracket P \rrbracket,\bisimd(\sigma_1,\sigma_2))$ (no over-approximation). As demonstrated in Example~\ref{prop:over_upper_bound_dda}, the approximation functionals would require for process variables $x \in \VarTerm(t)$ besides the bisimulation distance between $\sigma_1(x)$ and $\sigma_2(x)$ also information about the reactive behavior and the branching. However, for our objective to study the distance of composed processes in relation to the distance of its components, the bisimulation distance is the right level of abstraction.

We introduce now an order on $\npmultSym$ that ensures monotonicity of both the approximation functional $\da$ and the functional $\approxFunctor$ introduced in the next section to compute the denotation of arbitrary terms of a PGSOS PTSS. The order is defined by
\[
	P_1 \ordSPMultSym P_2 \text{ iff for all } p_1 \in P_1 \text{ there is a } p_2 \in P_2 \text{ such that } p_1 \ordPMultSym p_2. 
\]

\begin{proposition} \label{prop:preorder_D}
$(\npmultSym,\sqsubseteq)$ is a complete lattice.
\end{proposition}

We order process distances by $e_1 \sqsubseteq e_2$ iff $e_1(x) \le e_2(x)$ for all $x \in \Var$. The nondeterministic probabilistic distance approximation $\da$ is monotone in both arguments.

\begin{proposition} \label{prop:monotonicity_da}
Let $P,P' \in \npmultSym$ and $e,e' \in \distSym$. Then $\da(P,e) \le \da(P',e)$ if $P \sqsubseteq P'$, and $\da(P,e) \le \da(P,e')$ if $e \sqsubseteq e'$.
\end{proposition}

We will see in the following section that the denotations developed for terms of $P_{\text{PA}}$ are sufficient for terms of any PGSOS PTSS.

\section{Distance of composed processes}\label{sec:distance_composed_processes}

Now we provide a method to determine the denotation of an arbitrary term. In line with the former section this gives an upper bound on the bisimulation distance of closed instances of that term. In particular, the denotation for the term $f(x_1,\ldots,x_{\rank(f)})$ gives an upper bound on the distance of processes composed by the process combinator $f$. This allows us in the next section to formulate a simple condition to decide if a process combinator is uniformly continuous, and hence if we can reason compositionally over processes combined by that process combinator.

\subsection{Operations on process denotations}
We start by defining two operations on process denotations that allow us to compute the denotation of process terms by induction over the term structure. We define the operations first on $\multSym$ and then lift them to $\npmultSym$. 

The composition of two processes $t_1$ and $t_2$ which both proceed requires that their multiplicities are summed up (cf.\ parallel composition in the prior section). We define the \emph{summation} of multiplicities by:
\begin{align*}
	(m_1 \oplus m_2)(x) &= m_1(x) + m_2(x) 
\end{align*}
In order to define by structural induction the multiplicity of a term $f(t_1,\ldots,t_{\rank(f)})$, we need an operation that composes the multiplicity denoting the operator $f$ with the multiplicity of $t_i$. We define the \emph{pointed multiplication} of multiplicities with respect to variable $y \in \Var$ by:
\begin{align*}
	(m_1 \odot_{y} m_2)(x) &= m_1(y) \cdot m_2(x) 
\end{align*}
Then, the multiplicity of a state term $f(t_1,\ldots,t_{\rank(f)})$ is given by:
\[
	\llbracket f(t_1,\ldots,t_{\rank(f)}) \rrbracket_\multSym 
	= 
	\bigoplus_{i=1}^{\rank(f)} \left(\llbracket f(x_1,\dots,x_{\rank(f)}) \rrbracket_\multSym \odot_{x_i} \llbracket t_i \rrbracket_\multSym \right)
\]

\begin{example}
Consider the open term $t = a.x \parallel y$. From Section~\ref{sec:denotational_model_PA} we get $\llbracket a.x \rrbracket_\multSym = 1_x$, $\llbracket y \rrbracket_\multSym = 1_y$ and $\llbracket x_1 \parallel x_2 \rrbracket_\multSym = 1_{\{x_1,x_2\}}$. Then, we have $\llbracket t \rrbracket_\multSym =
(\llbracket x_1 \parallel x_2 \rrbracket_\multSym \odot_{x_1} \llbracket a.x \rrbracket_\multSym) \oplus 
(\llbracket x_1 \parallel x_2 \rrbracket_\multSym \odot_{x_2} \llbracket y \rrbracket_\multSym) = 
((1_{x_1} \oplus 1_{x_2}) \odot_{x_1} 1_x) \oplus ((1_{x_1} \oplus 1_{x_2}) \odot_{x_2} 1_y) = 
1_{\{x,y\}}$. 
\end{example}

It remains to define the multiplicity of $f(x_1,\dots,x_{\rank(f)})$ for operators $f$ with an operational semantics defined by some rule $r$. 
We define the multiplicity of $f(x_1,\dots,x_{\rank(f)})$ in terms of the multiplicity of the target of $r$. Let $\mu$ be a derivative of the source variable $x$ in rule $r$. We use the property $(m \odot_{\mu} 1_x)(x) = m(\mu)$ in order to express the multiplicity $m(\mu)$ as a multiplicity of $x$. Then, the multiplicity of $f(x_1,\dots,x_{\rank(f)})$ is defined for any variable $x$ as the summation of the multiplicity of $x$ and its derivatives in the rule target:
\[
	\llbracket \target(r) \rrbracket_\multSym \oplus \Biggl(\bigoplus_{x_i \trans[a_{i,m}] \mu_{i,m} \in \atop \pprem{r}}\llbracket \target(r) \rrbracket_\multSym \odot_{\mu_{i,m}} 1_{x_i}\Biggr)
\]

\begin{example} \label{ex:denotation_rule_without_testing}
Consider $t=f(x)$ and the following rule $r$:
\[ 
	\SOSrule{x \trans[a] \mu}{f(x) \trans[a] \mu \parallel \mu}
\]
The operator $f$ mimics the action $a$ of its argument, replicates the derivative $\mu$, and proceeds as a process that runs two instances of the derivative in parallel. Consider again the closed substitutions $\sigma_1(x)=a.a.0$ and $\sigma_2(x)=a.([0.9]a.0 \oplus [0.1]0)$ with  $\bisimd(\sigma_1(x),\sigma_2(x)) = 0.1$. Then, $\bisimd(\sigma_1(t),\sigma_2(t))=1-(1-0.1)^2$. The denotation of the target of $r$ is $\llbracket \target(r) \rrbracket_\multSym = 2_\mu$. Hence, the denotation of
$t$ is $2_\mu \oplus (2_\mu \odot_\mu 1_{x}) = 2_{\{\mu,x\}}$. Thus, $\dda(\llbracket t \rrbracket_\multSym,\bisimd(\sigma_1,\sigma_2))=1-(1-0.1)^2$ by 
$\bisimd(\sigma_1,\sigma_2)(x)=0.1$ and $\bisimd(\sigma_1,\sigma_2)(\mu)=0$.
\end{example}
Operations $\mathit{op} \in \{\oplus,\odot_{y}\}$ over $\multSym$ lift to $\npmultSym$ by
\begin{align*}
	(p_1  \; \mathit{op}  \; p_2)(m) &= \sum_{m_1,m_2 \in \multSym \atop m=m_1 \; \mathit{op} \; m_2} p_1(m_1) \cdot p_2(m_2) \\
	p \in (P_1  \; \mathit{op}  \; P_2) & \text{ iff } \exists p_1 \in P \text{ and } p_2 \in P_2 \text{ such that }p \sqsubseteq  p_1  \; \mathit{op}  \; p_2 \\
\end{align*}
%

\subsection{Approximating the distance of composed processes}

Let $(\Sigma, \Act, R)$ be any PGSOS PTSS. We compute the denotation of terms and rules as least fixed point of a monotone function. Let $S = S_T \times S_R$ with $S_T = \openTerms \to \npmultSym$ and $S_R = R \to  \npmultSym$. A pair $(\tau,\rho) \in S$ assigns to each term $t \in \openTerms$ its denotation $\tau(t) \in \npmultSym$ and to each rule $r \in R$ its denotation $\rho(r) \in \npmultSym$. Let $\mathcal{S} = (S,\sqsubseteq)$ be a poset with ordering $(\tau,\rho) \sqsubseteq (\tau',\rho')$ iff $\tau(t) \sqsubseteq \tau'(t)$ and $\rho(r) \sqsubseteq \rho'(r)$ for all $t \in \openTerms$ and $r \in R$. $\mathcal{S}$ forms a complete lattice with least element $(\bot_T,\bot_R)$ defined by $\bot_T(t) = \bot_R(r) = 0$ for all $t \in \openTerms$ and $r \in R$.

\begin{proposition} \label{prop:S_preorder}
$\mathcal{S}$ is a complete lattice.
\end{proposition}

We assume that for all rules $r \in R$ the source variable of argument $i$ is called $x_i$. Let $X_r$ be the set of source variables $x_i$ for which $r$ tests the reactive behavior, i.e. $x_i \in X_r$ iff $r$ has either some positive premise $x_i \trans[a_{i,m}] \mu_{i,m}$ or some negative premise $x_i \ntrans[b_{i,n}]$. 

The mapping $\approxFunctor \colon S \to S$ defined in Figure~\ref{fig:functor_multiplicity_metric_bisim} computes iteratively the nondeterministic probabilistic multiplicities for all terms and rules. As expected, the denotation of a state term $f(t_1,\dots, t_{\rank(f)})$ is defined as the application of all rules $R_f$ to the denotation of the arguments. However, for distribution terms the application of the operator needs to consider two peculiarities. First, different states in the support of a distribution term $f(\theta_1,\ldots,\theta_{\rank(f)})$ may evolve according to different rules of $R_f$.

\begin{figure}[!t]
\begin{empheq}[box=\widefbox]{align*}
\\[-0.1cm]
&\!\!\!\!\!\!\!\!\!\!\!\text{Function }\approxFunctor\colon S \to S \text{ is defined by }\approxFunctor(\termda,\ruleda) = (\termda',\ruleda') \text{ with}\\[-0.05cm]
	\termda'(t) &= 
	\begin{cases}
		1_{x} & \text{if } t=x \\[0cm]
		\displaystyle \bigoplus_{i=1}^{\rank(f)} \bigl(\rho_f \odot_{x_i} \termda(t_i) \bigr) \qquad\qquad & \text{if }
		\left[
		\begin{array}{l}	
			t=f(t_1,\ldots,t_{\rank(f)}) \\
			\rho_f =  \displaystyle\bigcup_{r \in R_f} \rho(r)
		\end{array} \right. \\[0.5cm]
	\end{cases}\\[0.05cm]
    \termda'(\theta) &= 
	\begin{cases}
 		1_\mu	& \text{if } \theta=\mu \\[0.2cm]
		\termda(t) 	& \text{if } \theta = \delta(t) \\[0.2cm]
 		\displaystyle \sum_{i\in I} q_i \cdot \termda(\theta_i) & \text{if } \displaystyle \theta=\sum_{i\in I} q_i \theta_i \\[0.5cm]
		\displaystyle \bigoplus_{i=1}^{\rank(f)} \bigl( \rho_f \odot_{x_i} \termda(\theta_i) \bigr) \qquad\quad & \text{if } 
		\left[
		\begin{array}{l}	
			\theta=f(\theta_1,\ldots,\theta_{\rank(f)}) \\
			\rho_f =  \displaystyle \downset{\left\{ \sup_{r \in R_f} \sup(\sup \rho(r),1_{X_r}) \right\}}
		\end{array} \right. \\[0.5cm]
	\end{cases}\\[0.1cm]
	\ruleda'(r) &= \Biggl\{ {p \oplus \Bigl( \bigoplus_{x_i \trans[a_{i,m}] \mu_{i,m} \in \atop \pprem{r}} p \odot_{\mu_{i,m}} 1_{x_i} \Bigr)} \mid p \in \termda(\target(r)) \Biggr\}
\end{empheq}
\caption{Computation of the denotation of arbitrary terms}
\label{fig:functor_multiplicity_metric_bisim}
\end{figure}

\begin{example} \label{ex:nondet_choice_on_distribution}
Consider the operator $f$ defined by the following rule:
\[
	\SOSrule{x \trans[a] \mu}{f(x) \trans[a] \mu + \mu}
\]
Operator $f$ replicates the derivative of $x$  and evolves as alternative composition of both process copies. Consider the closed substitutions $\sigma_1(x) = a.([0.9] a.a.0 \oplus [0.1] 0)$ and $\sigma_2(x) = a.([0.9] a.0 \oplus  [0.1] 0)$ with $\bisimd(\sigma_1(x),\sigma_2(x)) = 0.9$. Then, $\bisimd(\sigma_1(f(x)),\sigma_2(f(x))) = 1-0.1^2=0.99$. The denotations for the two rules defining the alternative composition (see Section~\ref{sec:denotational_model_PA}) are the downward closed sets with maximal elements $1_{\{x_1,\mu_1\}}$ and $1_{\{x_2,\mu_2\}}$. Since $\sup 1_{\{x_1,\mu_1\}} = 1_{\{x_1,\mu_1\}} \in \pmultSym$, $\sup 1_{\{x_2,\mu_2\}} = 1_{\{x_2,\mu_2\}} \in \pmultSym$ and $1_{X_{r^+_1}}=\{ x_1 \}$, $1_{X_{r^+_2}}=\{ x_2 \}$ we get 
$\rho_+ = \downset{\{{\sup(1_{\{x_1,\mu_1\}},1_{\{x_2,\mu_2\}})}\}} = {1_{\{x_1,x_2,\mu_1,\mu_2\}}} \in \npmultSym$. Hence, the denotation for the target of the $f$-defining rule is 
$\llbracket \mu + \mu \rrbracket = 
(1_{\{x_1,x_2,\mu_1,\mu_2\}} \odot_{x_1} 1_\mu) \oplus (1_{\{x_1,x_2,\mu_1,\mu_2\}} \odot_{x_2} 1_\mu) =
2_\mu$. Thus, $\llbracket f(x)\rrbracket = 2_x$. Then, $\dda(2_x,\bisimd(\sigma_1,\sigma_2)) = 
0.99$.
\end{example}

Second, in the distribution term $f(\theta_1,\ldots,\theta_{\rank(f)})$ the operator $f$ may discriminate states in derivatives belonging to $\theta_i$ solely on the basis that in some rule $r \in R_f$ the argument $x_i \in X_r$ gets tested on the ability to perform or not perform some action.

\begin{example} \label{ex:reactive_testing_of_distribution}
Consider the operators $f$ and $g$ defined by the following rules:
\begin{equation*}
	\SOSrule{x \trans[a] \mu}{f(x) \trans[a] g(\mu)}
\qquad\qquad
	\SOSrule{y \trans[a] \mu'}{g(y) \trans[a] \delta(0)} 
\end{equation*}
Operator $f$ mimics the first move of its argument and then, by operator $g$, only tests the states in the derivative for their ability to perform action $a$. Consider first operator $g$. We get $\bisimd(\sigma_1(g(y)),\sigma_2(g(y)))=0$ for all closed substitutions $\sigma_1,\sigma_2$.
Clearly, $\llbracket g(y) \rrbracket = \nullDPM$. Consider now $t=f(x)$ and substitutions $\sigma_1(x)=a.a.0$ and $\sigma_2(x)=a.([0.9]a.0 \oplus [0.1]0)$ with $\bisimd(\sigma_1(x),\sigma_2(x))=0.1$. The distance between $\sigma_1(f(x))$ and $\sigma_2(f(x))$ is the distance between distributions $\delta_{g(a.0)}$ and $0.9\delta_{g(a.0)} + 0.1\delta_{g(0)}$. From $\bisimd(g(a.0),g(0))=1$ we get $\bisimd(f(\sigma_1(x)),f(\sigma_2(x)))$ =
$\Kantorovich(\bisimd)(\delta_{g(a.0)},0.9\delta_{g(a.0)} + 0.1\delta_{g(0)})=0.1$. 

If we would ignore that $g$ tests its argument on the reactive behavior, then the denotation of $g(\mu)$ would be $\llbracket g(\mu) \rrbracket =  \llbracket g(x) \rrbracket \odot_{x} 1_\mu = 0$, and the denotation of $f(x)$ would be $\llbracket g(\mu) \rrbracket  \oplus (\llbracket g(\mu) \rrbracket \odot_{\mu} 1_x) = 0$. Then $\dda(0,\bisimd(\sigma_1,\sigma_2)) = 0 < 0.1 = \bisimd(f(\sigma_1(x)),f(\sigma_2(x)))$. 

Because the operator $g$ tests its argument on the ability to perform action $a$, it can discriminate instances of the derivative $\mu$ the same way as if the process would progress (without replication).  
Thus, the denotation of operator $g$ if applied in the rule target is 
$\rho_g = \downset{\{\sup(\sup 0,1_{X_{r^g}})\}} = 1_x$ as $X_{r^g}=\{x\}$. 
Hence, $\llbracket g(\mu) \rrbracket =  \rho_g \odot_{x} 1_\mu = 1_\mu$. Thus, $\llbracket f(x) \rrbracket = 1_x$.
It follows, $\bisimd(f(\sigma_1(x)),f(\sigma_2(x))) \le \dda(\llbracket f(x) \rrbracket,\bisimd(\sigma_1,\sigma_2)) = 0.1$.
\end{example}
\noindent
To summarize Examples~\ref{ex:nondet_choice_on_distribution} and~\ref{ex:reactive_testing_of_distribution}: The nondeterministic probabilistic multiplicity for operator $f$ applied to some distribution term is given by $\rho_f =  \downset{\{ \sup_{r \in R_f} \sup(\sup \rho(r),1_{X_r})) \}}$ (Figure~\ref{fig:functor_multiplicity_metric_bisim}). We explain this expression stepwise. For any rule $r$ we define by $\sup \rho(r) \in \pmultSym$ the least probabilistic multiplicity which covers all nondeterministic choices represented by the probabilistic multiplicities in $\rho(r) \in \npmultSym$. By $\sup(\sup \rho(r),1_{X_r}) \in \pmultSym$ we capture the case that premises of $r$ only test source variables in $X_r$ on their ability to perform an action (Example~\ref{ex:reactive_testing_of_distribution}). By $\sup_{r \in R_f} \sup(\sup \rho(r),1_{X_r}) \in \pmultSym$ we define the least probabilistic multiplicity which covers all choices of rules $r \in R_f$ (Example~\ref{ex:nondet_choice_on_distribution}). Finally, by the downward closure 
$\downset{\{ \sup_{r \in R_f} \sup(\sup \rho(r),1_{X_r})) \}} \in \npmultSym$ we gain the nondeterministic probabilistic multiplicity $\rho_f$ that can be applied to 
the distribution term (Figure~\ref{fig:functor_multiplicity_metric_bisim}).

\begin{proposition} \label{prop:F_monotone_and_continuous}
$\approxFunctor$ is order-preserving and upward $\omega$-continuous.
\end{proposition}
From Proposition~\ref{prop:S_preorder} and~\ref{prop:F_monotone_and_continuous} and the Knaster-Tarski fixed point theorem we derive the existence and uniqueness of the least fixed point of $\approxFunctor$. We denote by $(\lfp_T,\lfp_R)$ the least fixed point of $\approxFunctor$. We write $\llbracket t \rrbracket$ for $\lfp_T(t)$ and $\llbracket t \rrbracket_\tau$ for $\tau(t)$. We call $\llbracket t \rrbracket$ the canonical denotation of $t$. It is not hard to verify that all denotations presented in Section~\ref{sec:denotational_model_PA} for $P_{\text{PA}}$ are canonical.

A denotation of terms $\tau \in S_T$ is \emph{compatible} with a distance function $d \in [0, 1]^{\closedSTerms \times \closedSTerms}$, notation $d \preceq \llbracket \cdot \rrbracket_\tau$, if $d(\sigma_1(t),\sigma_2(t)) \le \da(\llbracket t \rrbracket_{\termda}, d(\sigma_1,\sigma_2))$ 
for all $t \in \openSTerms$ and all closed substitutions $\sigma_1,\sigma_2$. Now we can show that the functional $\Bisimulation$ to compute the bisimulation distance and functional $\approxFunctor$ to compute the denotations preserve compatibility (Proposition~\ref{prop:D_and_F_distribute_over_P}). A simple inductive argument allows then to show that the canonical denotation of terms $\llbracket \cdot \rrbracket$ is compatible with the bisimilarity metric $\bisimmetric$ (Theorem~\ref{thm:bounded_expansion_metric_bisim}). 


\begin{proposition} \label{prop:D_and_F_distribute_over_P}
Let $d \in [0, 1]^{\closedSTerms \times \closedSTerms}$ with $d \sqsubseteq \Bisimulation(d)=d'$ and $(\termda,\ruleda) \in S$ with $(\termda,\ruleda) \sqsubseteq \approxFunctor(\termda,\ruleda) = (\termda',\ruleda')$. Then $d \preceq \llbracket \cdot \rrbracket_{\termda} \text{ implies } d' \preceq \llbracket \cdot \rrbracket_{\termda'}$.
\end{proposition}

\begin{theorem}\label{thm:bounded_expansion_metric_bisim}
Let $P$ be any PGSOS PTSS with $\bisimmetric$ the bisimilarity metric on the associated PTS and $\llbracket \cdot \rrbracket$ the canonical denotation of terms according to $P$. Then $\bisimmetric \preceq \llbracket \cdot \rrbracket$.
\end{theorem}
\emph{Proof sketch.}
Remind that $\bisimd$ is the least fixed point of $\Bisimulation \colon [0,1]^{\closedSTerms \times \closedSTerms} \to [0,1]^{\closedSTerms \times \closedSTerms}$ defined by $\Bisimulation(d)(t,t') = \sup_{a\in A} \left\{ \Hausdorff(\Kantorovich(d))(\mathit{der}(t,a), \mathit{der}(t',a)) \right\}$ and $\Hausdorff$ the Hausdorff metric functional (Proposition~\ref{prop:bisim_metric_lfp_D}). 
Let $d_n=\Bisimulation^n(\botd)$ and $(\termda_n,\ruleda_n)=\approxFunctor^n(\bot_T,\bot_R)$. Proposition~\ref{prop:D_and_F_distribute_over_P} shows that $d_n \preceq \llbracket \cdot \rrbracket_{\termda_n}$ by reasoning inductively over the transitions specified by the rules. Monotonicity and upward $\omega$-continuity (Proposition~\ref{prop:F_monotone_and_continuous}) ensures that this property is also preserved in the limit.
\qed

\section{Compositional Reasoning} \label{sec:specification_compositional_process_combinators}


In order to reason compositionally over probabilistic processes it is enough if the distance of the composed processes can be related to the distance of their parts. This property is known as uniform continuity. In essence, compositional reasoning over probabilistic processes is possible whenever a small variance in the behavior of the parts leads to a bounded small variance in the behavior of the composed processes. Technically this boils down to the existence of a modulus of continuity. Uniform continuity generalizes earlier proposals of non-expansiveness~\cite{DGJP04} and non-extensiveness~\cite{BBLM13b}.

\begin{definition}[Modulus of continuity]
Let $f \in \Sigma$ be any process combinator. A mapping $z \colon [0,1]^{\rank(f)} \to [0,1]$ is a modulus of continuity for operator $f$ if $z(0,\ldots,0)=0$, $z$ is continuous at $(0,\ldots,0)$, and
\[
	\bisimd(f(t_1,\dots,t_{\rank(f)}), f(t_1',\dots,t_{\rank(f)}')) 
		\le
	z(\bisimd(t_1,t'_1),\ldots,\bisimd(t_{\rank(f)},t'_{\rank(f)}))
\]
for all closed terms $t_i,t_i' \in \closedSTerms$.
\end{definition}

\begin{definition}[Uniformly continuous operator]\label{def:uniform-continuity:combinator}
A process combinator $f \in \Sigma$ is \emph{uniformly continuous} if $f$ admits a modulus of continuity.
\end{definition}

Intuitively, a continuous binary operator $f$ ensures that for any non-zero bisimulation distance $\epsilon$ (understood as the admissible tolerance from the operational behavior of the composed process $f(t_1,t_2)$) there are non-zero bisimulation distances $\delta_1$ and $\delta_2$ (understood as the admissible tolerances from the operational behavior of the processes $t_1$ and $t_2$, respectively) such that the distance between the composed processes $f(t_1,t_2)$ and $f(t_1',t_2')$ is at most $\epsilon=z(\delta_1,\delta_2)$ whenever the component $t_1'$ (resp.\ $t_2'$) is in distance of at most $\delta_1$ from $t_1$ (resp.\ at most $\delta_2$ from $t_2$). 
We consider the uniform notion of continuity because we aim for universal compositionality guarantees. 

The denotation of $f(x_1,\ldots,x_{\rank(f)})$ allows to derive a candidate for the modulus of continuity for operator $f$ as follows.
\begin{definition}[Derived modulus of continuity]
Let $P$ be any PGSOS PTSS. For any operator $f \in \Sigma$ we define
\[
	z_f(\epsilon_1,\ldots,\epsilon_{\rank(f)}) = \min\left( \sum_{i=1}^{\rank(f)} m_f(x_i) \epsilon_i , 1 \right)
\]
with $m_f = \overline{\sup \llbracket f(x_1,\dots,x_{\rank(f)}) \rrbracket}$. 
\end{definition}

Trivially, we have $z_f(0,\ldots,0)=0$ and $\bisimd(f(t_1,\dots,t_{\rank(f)}),f(t_1',\dots,t_{\rank(f)}')) \le z_f(\bisimd(t_1,t'_1),\ldots,\bisimd(t_{\rank(f)},t'_{\rank(f)}))$ for all closed terms $t_i,t_i' \in \closedSTerms$ by Theorem~\ref{thm:bounded_expansion_metric_bisim}. However, $z_f$ is continuous at $(0,\ldots,0)$ only if the multiplicities in the denotation $\llbracket f(x_1,\ldots,x_{\rank(f)} \rrbracket$ assign to each variable a finite value.

\begin{theorem} \label{thm:uniform_ops}
Let $P$ be any PGSOS PTSS. A process combinator $f \in \Sigma$ is uniformly continuous if
\[
	\llbracket f(x_1,\dots,x_{\rank(f)}) \rrbracket \sqsubseteq n_{\{x_1,\ldots,x_{\rank(f)}\}}
\]
for some $n \in N$.
\end{theorem}

\begin{example}
We will show that unbounded recursion operators may be not uniformly continuous. We consider the replication operator of $\pi$-calculus specified by the rule:
\begin{equation*}
	\SOSrule{x \trans[a] \mu}{!x \trans[a] \mu \parallel \delta(!x)}
\end{equation*}

The 
replication operator is not continuous since no $z$ with $\bisimd(!t,!t') \le z(\bisimd(t,t'))$ and $z(0)=0$ will be continuous at $0$ since $z(\delta)=1$ for any $\delta>0$. The denotation $\llbracket !x \rrbracket = \infty_x$ shows that the argument $x$ is infinitely often replicated. Hence, the replication operator is not continuous.

\end{example}

Even more, for uniformly continuous operators $f$ the function $z_f$ is a modulus of continuity. 

\begin{theorem} \label{thm:modulus_continuity_of_op}
Let $P$ be any PGSOS PTSS. A uniformly continuous process combinator $f \in \Sigma$ satisfies
\[
	\bisimd(f(t_1,\dots,t_{\rank(f)}), f(t_1',\dots,t_{\rank(f)}')) 
		\le
	z_f(\bisimd(t_1,t'_1),\ldots,\bisimd(t_{\rank(f)},t'_{\rank(f)}))
\]
for all closed terms $t_i,t_i' \in \closedSTerms$.
\end{theorem}

In reverse, for a given modulus of continuity (as specification of some process combinator), we can derive the maximal replication of process of this operator.

\begin{definition}[Derived multiplicity]
Let $z \colon [0,1]^n \to [0,1]$ be a mapping with $z(0,\ldots,0)=0$ and $z$ continuous at $(0,\ldots,0)$. Let $m \colon \Var \to \Rgezinfty$ be defined by
\[
	m = \sup \left\{ m \colon \Var \to \Rgezinfty \mid \forall e \in \distSym.\, \sum_{i=1}^{n} m(x_i) e(x_i) \le z(e(x_1),\ldots,e(x_n)) \right\}
\]
where $m_1,m_2 \colon \Var \to \Rgezinfty$ are ordered $m_1 \sqsubseteq m_2$ iff $m_1(x) \le m_2(x)$ for all $x \in \Var$.
We call $m$ the derived multiplicity of $z$.
\end{definition}
\begin{theorem} \label{thm:op_for_modulus_continuity}
Let $P$ be any PGSOS PTSS, $z \colon [0,1]^n \to [0,1]$ be a mapping with $z(0,\ldots,0)=0$ and $z$ continuous at $(0,\ldots,0)$, and $m$ the derived multiplicity of $z$. Then, an operator $f \in \Sigma$ with $\rank(f)=n$ has $z$ as modulus of continuity if
\[
	\overline{\sup \llbracket f(x_1,\dots,x_{\rank(f)}) \rrbracket} \sqsubseteq m
\]
\end{theorem}

To conclude, the methods provided in Section~\ref{sec:denotational_model_PA} and~\ref{sec:distance_composed_processes} to compute an upper bound on the distance between instances of the term $f(x_1,\ldots,x_{\rank(f)})$ can be used to derive the individual compositionality property of operator $f$ given by its the modulus of continuity $z_f$. 
Note that $z_f$ depends on all those rules which define operators of processes to which an instance of $f(x_1,\ldots,x_{\rank(f)})$ may evolve to. Traditional rule formats define syntactic criteria on single rules in order to guarantee a desired compositionality property of the specified operator. In contrast, our approach derives the compositionality property of an operator from the the syntactic properties of those rules which define the operational behavior of processes composed by that operator.
\section{Conclusion and Future Work}\label{sec:conclusion}

We presented a method to approximate the bisimulation distance of arbitrary process terms (Theorem~\ref{thm:upper_bound_da} and~\ref{thm:bounded_expansion_metric_bisim}). This allows to decide for any given PTSS which operators allow for compositional metric reasoning, i.e. which operators are uniformly continuous (Theorem~\ref{thm:uniform_ops}). 
Moreover, our method allows to compute for any given PTSS a modulus of continuity of each uniformly continuous operator (Theorem~\ref{thm:modulus_continuity_of_op}). Additionally, for any given modulus of continuity (understood as the required compositionality property of some operator) we provide a sufficient condition to decide if an operator satisfies the modulus of continuity (Theorem~\ref{thm:op_for_modulus_continuity}). The condition characterizes the maximal number of times that processes combined by the operator may be replicated during their evolution in order to guarantee the modulus of continuity.

We will continue this line of research as follows. First, we will investigate the compositionality of process combinators with respect to convex bisimulation metric~\cite{DAMRS07}, discounted bisimulation metric~\cite{DGJP04}, and generalized bisimulation metric~\cite{CGPX14}.
Second, we will explore compositionality with respect to behavioral pseudometrics based on trace semantics~\cite{AFS04} and testing semantics. Finally, we will investigate how the denotational approach to decide the compositionality properties of operators relates to the logical approach to derive rule formats of~\cite{BFvG04,GF10}. Besides this general line, we want to investigate how our structural syntactic approach to compositionality relates to the algorithmic computational approach in~\cite{BBLM13b}.


\bibliographystyle{eptcs}
\bibliography{paper}



\end{document}